\pdfoutput=1
\documentclass[aps,prd,10pt,amsmath,floats,floatfix, twocolumn,superscriptaddress,nofootinbib,showpacs,longbibliography]{revtex4-1}
\usepackage{titlesec,titleps,titletoc}
\usepackage[normalem]{ulem}
\usepackage{tocloft}
\usepackage{array}
\usepackage{xspace}
\usepackage[scr=boondoxo]{mathalpha}
\usepackage[T1]{fontenc}
\usepackage[utf8]{inputenc}
\usepackage[varg]{txfonts}

\usepackage{graphicx}
\usepackage{caption}
\usepackage{ragged2e}
\DeclareCaptionFormat{justified}{\justifying#1#2#3}
\usepackage{lipsum}
\usepackage{epstopdf}
\usepackage[mathscr]{euscript}
\usepackage{amssymb}
\usepackage{amsfonts}
\usepackage{amsthm}
\usepackage{mathtools}
\usepackage{color}
\usepackage{times}
\usepackage{verbatim}
\usepackage{lmodern}
\usepackage{url}
\usepackage{braket}
\usepackage{epsfig}
\usepackage{calc}
\usepackage{float}
\usepackage[
colorlinks=true,
linkcolor=blue,
urlcolor=magenta,
filecolor=blue,
citecolor=red,
pdfstartview=FitV,
pdftitle={},
pdfauthor={},
pdfsubject={},
pdfkeywords={},
pdfpagemode={},
bookmarksopen=true
]{hyperref}
\usepackage{cleveref}
\crefname{figure}{FIG.}{figures}
\Crefname{figure}{Figure}{Figures}
\newcommand{\ca}[1]{\mathcal{#1}}

\newcommand{\abs}[1]{\left|#1\right|}

\def\frac#1#2{{#1\over #2}}

\def\be{\begin{equation}}
	\def\ee{\end{equation}}
\def\ba{\begin{eqnarray}}
	\def\ea{\end{eqnarray}}

\newcommand{\red}{\color{red}}

\usepackage{microtype}
\usepackage[english]{babel}
\titleformat{\paragraph}[runin]{\normalfont\normalsize}{\paragraph}{1em}{}[. I\mbox{}]
\titlespacing*{\paragraph}{0pt}{3.25ex plus 1ex minus .2ex}{\the\fontdimen2\font}
\begin{document}
		\title{Can the Near-Horizon Black Hole Memory be detected through Binary Inspirals?}
    \author{Sajad A. Bhat}
    \email{sajad.bhat@iucaa.in}
    \affiliation{The Inter-University Centre for Astronomy and
    Astrophysics,\\ Post Bag 4, Ganeshkhind, Pune 411007, India }
    \author{Srijit Bhattacharjee}
    \email{srijuster@gmail.com}
    \thanks{The first two authors contributed equally}
    \affiliation{Indian Institute of Information Technology,  Allahabad,\\ Deoghat, Jhalwa, Prayagraj- 211015, India}
    \author{Shasvath J. Kapadia}
    \email{shasvath.kapadia@iucaa.in}
    \affiliation{The Inter-University Centre for Astronomy and
    Astrophysics,\\ Post Bag 4, Ganeshkhind, Pune 411007, India }
	\date{\today}
	\begin{abstract}
	The memory effect, in the context of gravitational-waves (GWs), manifests itself in the permanent relative displacement of test masses when they encounter the GWs. A number of works have explored the possibility of detecting the memory when the source and detector are separated by large distances. A special type of memory, arising from Bondi--Metzner--Sachs (BMS)-like symmetries, called ``black-hole memory'', has been recently proposed. The black hole memory only manifests itself in the vicinity of its event horizon. Therefore, formally observing it requires placing a GW detector at the horizon of the BH, which \textit{prima-facie} seems unfeasible. In this work, we describe toy models that suggest a possible way the black hole memory may be observed, without requiring a human-made detector near the event horizon. The models consider a binary black hole (BBH), emanating GWs observable at cosmological distances, as a proxy for an idealized detector in the vicinity of a supermassive black hole that is endowed with a supertranslation hair by sending a shock-wave to it. This sudden change affects the geometry near the horizon of the supertranslated black hole and it induces a change in the inspiraling orbital separation (and hence, orbital frequency) of the binary, which in turn imprints itself on the GWs. Using basic GW data analysis tools, we demonstrate that the black hole memory should be observable by a LISA-like space-based detector.
	\end{abstract}

	\maketitle
	
\section{Introduction}\label{introduction}

Testing different aspects of gravity has been an active field of research ever since the successful detection of gravitational waves (GWs) with the LIGO detectors in 2015 \cite{PhysRevLett.116.061102, PhysRevLett.125.101102}. One of the major issues in gravitational physics is to extract useful information from regions of spacetime where gravity is strong. Black holes are among the most exotic compact objects in the universe having strong gravitational effects near their horizons. GWs generated by violent astrophysical events like the collision of two black holes encode useful information regarding the physics near the horizons. 
 
The LIGO-Virgo (LV) detector network \cite{TheLIGOScientific:2014jea, TheVirgo:2014hva} has detected $\sim 90$ GW events produced by compact binary coalescences (CBCs), across three observing runs (O1, O2, O3) \cite{KAGRA:2021vkt}. The majority of these are binary black hole (BBH) mergers, and have therefore enabled some of the most unique tests of general relativity (GR) in the strong field regime \cite{LIGOScientific:2021sio}. These include a model-agnostic test of GR which subtracts out the best matched GR template from the data known to contain the GW signal, and examines if the residual is consistent with noise \cite{LIGOScientific:2016lio, LIGOScientific:2019fpa, LIGOScientific:2020ufj}; an inspiral-merger-ringdown consistency test that checks for consistency between the low and high frequency portions of the GW waveform \cite{Ghosh:2016qgn, Ghosh:2017gfp}; a test that looks for deviations in the Post-Newtonian (PN) parameters governing the inspiral evolution of the CBC \cite{blanchet1994signal, Blanchet:1994ez, Arun:2006hn, Arun:2006yw, Yunes:2009ke, Mishra:2010tp, Li:2011cg, Li:2011vx}; and propagation tests that compare the speed of GWs with the speed of light \cite{LIGOScientific:2018dkp}, as well as signatures of velocity dispersion due to a finite graviton mass \cite{Will:1997bb}.
 
 Gravitational memory effect has gained much attention in recent times due to multiple reasons. Memory effect can be understood as the permanent shift in the relative separation of two test detectors placed near null infinity after the passage of GWs \cite{1974SvA....18...17Z,1987Natur.327..123B}:
\[\Delta h_{+,\times}^{mem}=\lim_{t\to \infty}h_{+,\times}(t)-\lim_{t\to -\infty}h_{+,\times}(t)\]
This remarkable feature of spacetime is a genuine test of general relativity. The linear memory appears due to the gravitational radiation generated by binaries executing unbounded (hyperbolic) trajectories and is generally expected to be vanishing for binaries with bound orbits. \footnote{Although for bound orbits some contribution from linear, nonheriditary DC terms appears at 5PN and higher orders, it is not clear that these effects will lead to a permanent memory~\cite{Favata:2008yd}. Nevertheless, this effect is very weak and has a very low detection prospect.} However, there is another type of memory that originates at the 2.5PN order in the radiative multipoles  for quasi-circular and elliptical binary inspirals but contributes at the 0PN order in the waveform~\cite{Favata:2008yd}. This is in contrast to the hyperbolic, parabolic and radial orbits where this effect occurs at 2.5PN order relative to the leading order (Newtonian) term in the waveform~\cite{Favata:2011qi}. This is known as non-linear memory or Christodoulou memory \cite{1991PhRvL..67.1486C, PhysRevD.46.4304, PhysRevD.45.520}. This memory has a simple interpretation: it is the ``wave of a wave'', i.e, it arises due to GWs sourced by GWs. The detection prospects of nonlinear memory by ground-based and space-based detectors has been extensively studied~\cite{Favata:2008ti, Favata:2008yd, Favata:2009ii, Favata:2010zu, Favata:2011qi, Lasky:2016knh,Grant:2022bla, Ghosh:2023rbe}. This memory signal is expected to be detected by future space-based detectors like Laser Interferometer Space Antenna (LISA) (see, e.g., \cite{Favata:2009ii, Johnson:2018xly, Ghosh:2023rbe}). 

Recently, an interesting relation between the memory effect and asymptotic symmetries of spacetimes has been established. The asymptotic symmetries are the symmetries that preserve the asymptotic structure of spacetimes. This was first discovered by Bondi-van-der Burg-Metzner and Sachs (BMS) and described in their pioneering works \cite{doi:10.1098/rspa.1962.0161, PhysRev.128.2851}. The group of symmetries recovered near the infinity of an asymptotically flat spacetime is termed as BMS group. In its primitive form, the BMS group contained an infinite dimensional abelian subgroup whose generators are called supertranslations. These are angle dependent translations that generalize the ordinary rigid translations of flat spacetime. Later, the asymptotic group has been extended to accommodate superrotations and can be viewed as conformal symmetries of the celestial sphere at null infinity ($\ca{I}^{+}$) \cite{PhysRevLett.105.111103, barnich2012supertranslations}. 

Supertranslations have an interesting relation with the memory effect. Near null infinity, two test detectors would undergo a shift in their relative position due to the passage of GWs and this can be realized as an action of supertranslation on the boundary data (the physical content of metric at the boundary) \cite{PhysRevLett.116.231301, Strominger2016}. This intriguing fact indicates that the detection of the memory effect would be an alternative way to detect supertranslation like symmetries. Since supertranslation (or superrotations) are symmetries of the asymptotic solutions of gravitational theories, they would have their own charges. In fact, there will be an infinite number of conserved charges available due to the supertranslations (or superrotations). These charges reveal that the classical Minkowski vacuum is highly degenerate and different vacua are related by supertranslations \cite{Hawking2017, strominger2018lectures}. The gravitational memory is now described as a transition between two inequivalent vacua induced by a supertranslation. 

Such conservation laws should also be true in the presence of asymptotically flat black holes. However, in this case, the charge conservation requires the black holes should also carry supertranslation (or superrotations) charges or some (soft) `hairs'~\cite{Hawking2017, strominger2018lectures}.    
In the presence of black holes, the future null infinity ${\mathcal I}^{+}$ fails to act as a Cauchy surface. One needs to add the future horizon ${\mathcal H}^{+}$ also, and ${\mathcal I}^{+} \cup {\mathcal H}^{+}$ acts as a Cauchy surface.
Therefore, it is desirable to recover the BMS-like symmetries near the horizon of a black hole in spacetime as well. 

BMS-like symmetries have been recovered near the horizon of a black hole by several theoretical approaches. BMS-like symmetries can be discovered at the null boundaries including black hole horizons by analyzing diffeomorphisms that preserve certain universal structures of these surfaces \cite{Chandrasekaran2018,Ashtekar2018}. In another approach, near-horizon asymptotic symmetries are recovered by preserving the near-horizon asymptotic structures \cite{Koga_2001, PhysRevLett.116.091101, Donnay2016}\footnote{BMS-like symmetries have also been recovered as soldering symmetries of thin shells. Readers may look at \cite{Blau2016, PhysRevD.98.104009} and the references therein}. 

The emergence of BMS-like symmetries near the horizon of a black hole generates an obvious question: can one find a memory effect similar to what one recovers at null infinity? The answer to this question is affirmative.  Memory in the form of a shift in the vacua or solutions upon sending shock-wave on a black hole horizon has been described in \cite{PhysRevD.98.124016}. How near-horizon BMS memory can be imprinted on test observers falling freely near the horizon has also been described in \cite{Bhattacharjee2021}. If one considers the symmetries at the black hole horizon, then BMS memory can be defined as the change in certain parameters like expansion, shear related to the null congruences generating the horizon \cite{PhysRevD.101.124010, PhysRevD.100.084010,Adami:2021nnf,Sarkar:2021djs}.  

Clearly, the existence of such a near-horizon memory can only be detected by an idealized observer hovering around the horizon. Can such a memory be recognised from the far region? Particularly, would it be possible to think of a scenario in which the near-horizon memory can show up in future GW detectors? In this article, we propose a toy model that can provide plausible answers to these questions.

\section{Implantation of memory and its detection}
 
We consider two scenarios to detect BMS-like memories.  First, a simplified model where a bald Schwarzschild black hole is considered, and the implantation of supertranslation memory on it is indicated. Next, this simplified model is further improved by considering a rapidly spinning Kerr black hole endowed with a supertranslation hair. In both the cases we analyze the possible way of obtaining imprints of such near-horizon memory in GWs by placing an intermediate mass binary black hole (IMBBH) as proxies for idealized detectors. 

 \subsection{Memory for a non-rotating Black Hole}
 
 The detection setup uses a toy model in which an eternal Schwarzschild black hole is endowed with a linearised supertranslation hair by some asymmetric shockwave that fall into it as proposed in \cite{Hawking2017,strominger2018lectures}. This process implants a linearized supertranslation hair to a bald Schwarzschild black hole:
 \begin{align}\label{bald}
g_{\mu\nu}^{Schw}dx^{\mu}dx^{\nu}=-\left(1-\frac{2M}{r}\right)dv^2+r^2(d\theta^2+\sin^2\theta d\phi^2)\,.
 \end{align}
The metric with a supertranslation hair is produced by throwing an asymmetric pulse to the black hole at some advanced time $v_0$ from $\mathscr{I}^{-}$ with the following energy-momentum density:
 \begin{align}\label{Tab}
    T_{vv}=&\left(\frac{\mu+\hat{T}(\theta,\phi)}{4\pi r^2}+ \frac{D_A\hat{T}^A(\theta,\phi)}{4\pi r^3} \right)\delta(v-v_0),\\
    T_{vA}=&\frac{\hat{T}^A(\theta,\phi)}{4\pi r^2}\delta(v-v_0),
 \end{align}
where $D_A$ is the covariant derivative on the unit 2-sphere. $\hat{T}^A$ satisfies $(D^2+2)D_A\hat{T}^A=-6M\hat{T}$, where $D^2$ is the Laplacian on 2-sphere. The above equation can be solved using Green's function technique to yield the perturbed metric produced by the shock-wave profile \cite{Hawking2017,strominger2018lectures}. The effect of such a shock-wave is equivalent to the action of a (large) diffeomorphism on the background bald metric:
  \begin{align}\label{tem}
		h_{\mu \nu} = \left(\mathcal{L}_{\chi_{f}} g^{Schw}_{\mu\nu} + \dfrac{2 \mu}{r} \delta^{v}_{\mu} \delta^{v}_{\nu}\right)\Theta(v-v_0).
	\end{align}
The BMS Killing vector $\chi_f$ acts on the Schwarzschild metric and generates a supertranslated Schwarzschild black hole with a different mass parameter \cite{Hawking2017}. The function $f$ is related to the supertranslation symmetry. The shockwave acts as a boundary between a bald black hole and a black hole with supertranslation hair. In the supertranslated phase, $(v>v_0)$ the metric reads:
 \begin{eqnarray}\label{STSw}
 ds^2=-\left(1-\frac{2M}{r}-\frac{MD^2f}{r^2} \right)dv^2+2dvdr \nonumber\\-dv d\Theta^A D_A\left(2\left(1-\frac{2M}{r}\right) f+D^2f\right)+\nonumber\\\left( r^2\gamma_{AB}+2rD_AD_B f(\Theta)-r\gamma_{AB}D^2f \right)d\Theta^A d\Theta^B,
 \end{eqnarray}
 where $\Theta^A$ collectively denotes the angular coordinates $(\theta, \phi)$ on a sphere. This metric is exact in the radial coordinate $r$ but linear in the supertranslational field $f(\theta, \phi)$. The horizon of this supertranslated metric is at $r^h_{(f)}=2M+\frac{1}{2}D^2f.$ The transverse components of this metric undergo a shift due to the supertranslation that has been implanted to the horizon of the bald black hole:
 \begin{equation}\label{TC}
     \Delta h_{AB}=4M\left( D_AD_B f(\Theta)-\frac{1}{2}\gamma_{AB}D^2f \right).
 \end{equation}
 This change in the transverse part of the metric components gives rise to memory for an observer sitting at null infinity. As the physical degrees of freedom are encoded in the transverse part of the metric, this produces a genuine effect.  
 
BMS-like symmetries can also be recovered near the horizon of a black hole that preserves the near-horizon asymptotic structure \cite{Donnay2016,PhysRevD.98.124016}. To see this, one may expand the metric in Eq. (\ref{STSw}) near the horizon by a suitably defined radial coordinate $\rho=r-r^h_{(f)}$ and relate the parameters of the near-horizon metric with the implanted supertranslation field. To the leading order in $\rho$ this metric reads \cite{Donnay2016,PhysRevD.98.124016}:
\begin{align}\label{NHM1}
        ds^2= -&{\rho\over 2M}  dv^2+2dv d\rho -{\rho \over M} D_Af d\Theta^A dv \nonumber \\ 
        + &(4M^2\gamma_{AB} + 4M D_AD_B f)d\Theta^A d\Theta^B+\mathcal{O}(\rho^2)\,.
\end{align}
 It is not difficult to see that the transverse component of this metric to leading order is identical to (\ref{TC}) except for a term linear in $f$ absorbed in the definition of $\rho$. The bald counterpart of this metric would be obtained by setting $f=0$. 

 The effect of the near-horizon asymptotic Killing vector $\chi$ (eg. supertranslations) is to preserve the above metric (\ref{NHM1}) but change the background fields. For example, the transverse metric components $\Omega_{AB} $ undergo a change \cite{Donnay2016, PhysRevD.98.124016}:
 \begin{align}
     \delta_\chi \Omega_{AB}=4M D_AD_Bf.
 \end{align}
The changes induced via horizon supertranslation (or superrotation) are termed as `black hole memory effect' \cite{PhysRevD.98.124016}. The change induced by the shock-wave can be realized as the action of supertranslation and superroation on the asymptotic metric. From the perspective of a near-horizon observer the changes will get captured via the metric components \eqref{NHM1}. One can obtain the charges corresponding to these near-horizon symmetries. The supertranslation and superrotation charges depend upon the field $f$ or its derivatives, \footnote{see \cite{Donnay2016} for details.} and therefore, if we detect $f$ by any mechanism, we detect the supertranslation or the black hole memory. 

\subsection{A toy detection model}
\begin{figure*}[htb]
    \centering
    \includegraphics[width =\linewidth]{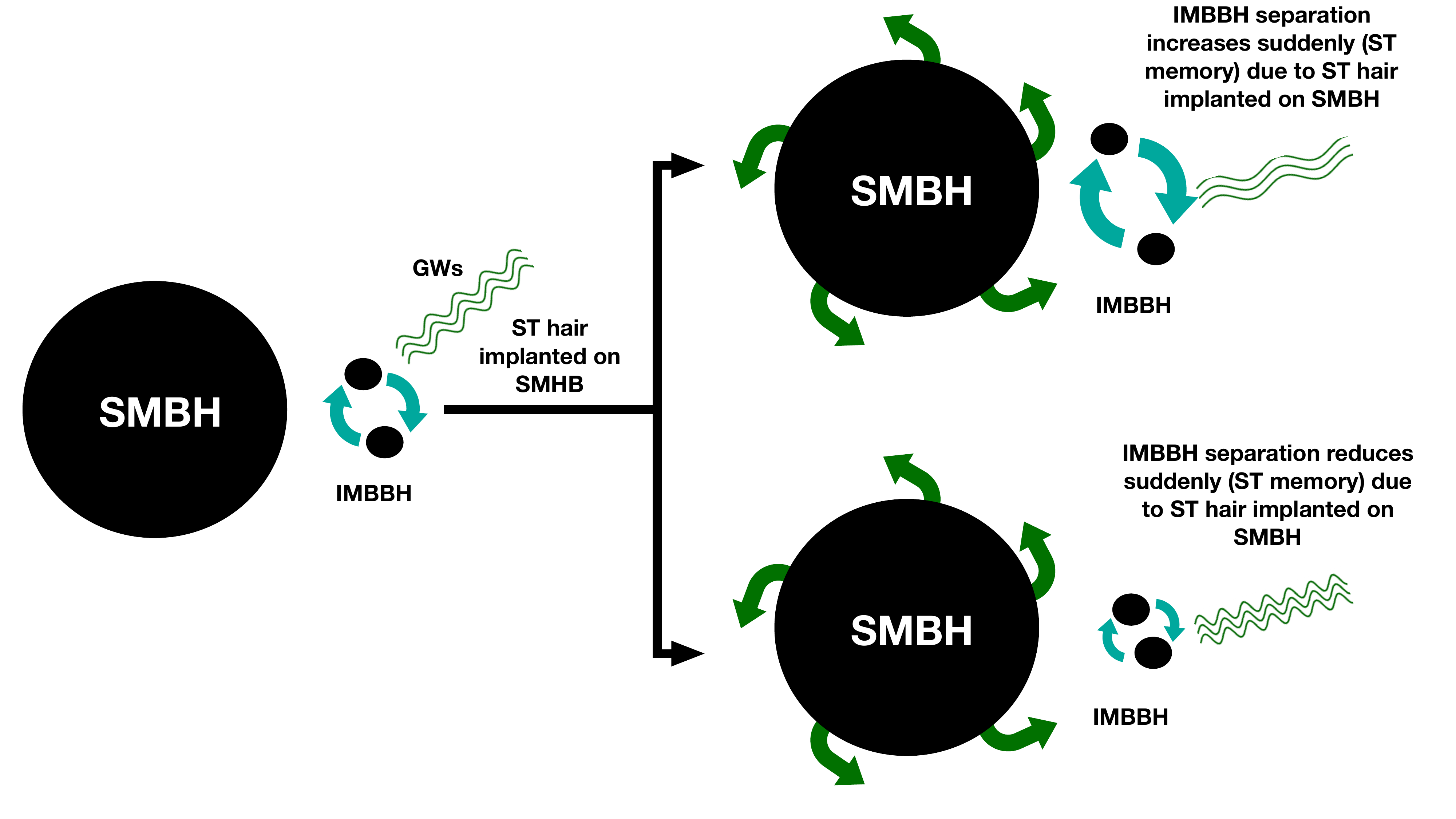}
  \caption{Schematic representation of inspiralling IMBBH source emitting GWs in the vicinity of the horizon of a bald SMBH (on the left). Once the supertranslation (ST) `hair' is attached to the bald SMBH due to some astrophysical shock wave, the proper distance between the components of the IMBBH is changed. The magnitude of the change in separation (ST memory) denoted by $\Delta L$ is given by Eqs. (~\ref{memory-sch},\ref{memory-kerr}). As the IMBBH separation increases (decreases), the binary will start emitting GWs at a lower (higher) frequency in comparison to the initial configuration (on the left).  } 
  \label{schematic-fig}
\end{figure*}
 It is apparent that to detect the near-horizon supertranslation memory, one needs to place the detectors close to the horizon. However, detection from a far region would not be possible until the detectors can have some mechanism to communicate with the remote region. Now, we propose a scenario (probe for detection) that might capture this supertranslation memory of a Schwarzschild black hole. Suppose that two intermediate-mass black holes (IMBHs) in a binary are in the vicinity of a bald supermassive black hole (SMBH) and they act like detectors. The IMBHs initially continue to generate GWs that are detectable by LISA. Now suddenly the SMBH (almost instantaneously) gets endowed with a (linear) supertranslation hair by sending a shockwave into it as described above. This sudden effect would change the relative proper distance between the inspiraling IMBHs and would ultimately alter the inspiraling frequency and frequency evolution as {\red illustrated in Figure~\ref{schematic-fig}.} Note that the shockwave acts only as an agent that implants the supertranslation hair on the bald SMBH. This is a convenient way to generate a BH state with a (asymptotic) supertranslation field. However, as already alluded, this changed metric can also be achieved via Lie transporting the bald metric.  
 
 We consider the IMBHs to be separated only in the angular directions initially, i.e. they are at the same radial distance from the centre of the SMBH. After the hair is implanted, the separation between the black holes changes. This can be estimated using local coordinates. We will further assume that the supertranslation field is a function of $\theta$ only. This would reduce the algebra and at the same time offer sufficient insight into the desired investigation.  
 
 Now, let $L_0$ be the initial proper distance (angular) between the test bodies (IMBHs). In that case, the separation would only be governed by the bald angular part of the metric. Therefore $L_0^2=4M^2 \gamma_{AB}d\Theta^A d\Theta^B.$ After the implantation of the supertranslation memory, the changed distance would be governed by the metric given by Eq.~(\ref{STSw}) or Eq.(\ref{NHM1}). If we consider the position of IMBHs almost near the horizon of SMBH, then we should go ahead with the metric in (\ref{NHM1}). However, this situation will practically be untractable from the far region, as the distance between the binary and the horizon of SMBH will be within the ISCO of SMBH. Therefore, we take the metric described in Eq.~(\ref{STSw}) which is valid at any finite distance from the black hole. If $\Delta L$ is the shift in the angular separation, then a straightforward computation yields:
          \begin{align}\label{memory1}
        \Delta L&\simeq \frac{r_0 \partial^2_{\theta}f(\delta\theta^2-\sin^2\theta \delta\phi^2) }{2L_0} \nonumber \\
       & \simeq -\frac{3r_0}{2L_0}q \cos(2\theta) (\delta\theta^2-\sin^2\theta \delta\phi^2),
        \end{align}
 where we have chosen $f(\theta)$ to be equal to the second degree Legendre Polynomial $f=qP_2(\cos \theta).$ $q$ is a parameter that encapsulates the strength of the supertranslation hair \cite{Sarkar:2021djs, Compere2016}. For the case under consideration, $q$ should be small, as it is part of a weak supertranslation field. For consistency, the initial separation $L_0$ should be much greater than the separation $\Delta L$ induced by the field $f$. Here, $\delta \theta$ and $\delta\phi$ are, respectively, the initial polar and azimuthal separations between the objects. The shift in the separation of IMBHs, parametrized by $q$, will induce a change in the frequency of their inspiraling motion. We set IMBBH at distance $r_0= (1+0.001)R$ with $R\simeq 3(2M+\frac{1}{2}D^2f)$ as the IMBHs should be at least in the vicinity of ISCO of SMBH. Further, we introduce a dimensionless parameter $q_0=q/M$ and rewrite the leading contribution from (\ref{memory1}) as
\begin{align}\label{memory-sch}
\Delta L\simeq -\frac{9M^2}{L_0}q_0 \cos(2\theta) (\delta\theta^2-\sin^2\theta \delta\phi^2).
\end{align}
We will use Eq.~ \eqref{memory-sch} to demonstrate the detectability of the mis-match in LISA between the original GW signal and the GW signal modulated due to this sudden change in the separation between the IMBHs in the section~\ref{freq-mod}.

\subsection{A more realistic model}
The model considered so far can be improved further. In the previous case, the ISCO of the background Schwarzschild SMBH is not very near to the horizon and one may wonder whether the detection of ST hair can be trusted as a genuine near-horizon memory. In this regard, we now consider a rapidly spinning Kerr black hole as the SMBH. A rapidly spinning Kerr black hole's metric can be presented as an extreme Kerr metric. Such rapidly spinning SMBHs are indeed expected in nature \cite{McClintock:2006xd, reynolds2015measuring, Gou:2013dna}. The extreme Kerr-metric near the horizon is obtained from the traditional Boyer-Lindquist (BL) form by employing several coordinate transformations \cite{Bardeen:1999px,Chrucel}. We start from the BL form:
\begin{equation}\label{eq:KerrBL}
    ds^2 = - ( \frac{d{\bar{r}^2}}{\bar{\Delta}} + {d\bar{\theta}^2} )  \bar{\rho}^2 + ( d{\bar{t}} - a \sin^2\bar{\theta} d{\bar{\phi}} )^2 \frac{\bar{\Delta}}{\bar{\rho}^2} - (\bar{A}^2 d{\bar{\phi}} - a  d{\bar{t}})^2 \frac{\sin^2 \bar{\theta}}{\bar{\rho}^2},
\end{equation}
with
\begin{align}
   & \bar{\Delta}(\bar{r}) = \bar{r}^2 + a^2 - 2M\bar{r}, \\
   & \bar{\rho}^2(\bar{r},\bar{\theta}) = \bar{r}^2 + a^2 \cos^2 \bar{\theta}, \\
   & \bar{A}^2(\bar{r}) = \bar{r^2} + a^2 ,
\end{align}
where $M, a$ are the mass and angular momentum per unit mass of the black hole. As depicted in \cite{Bardeen:1999px, Chrucel}, we produce a similar kind of metric as in Eq. (\ref{NHM1}) by performing the following transformations first: $\bar{t}=\epsilon^{-1} t', \bar{r}=M+\epsilon r',\bar{\theta}=\theta',\bar{\phi}=\phi'+\epsilon^{-1}\frac{a}{r_0^2}t'$, with $r_0^2=M^2+a^2,$ and taking $\epsilon \to 0$ limit, and then going to a regular advanced null coordinate compatible with the Bondi-Sachs gauge \cite{Fletcher-Lun}:
\begin{align}\label{nhek}ds^2=(1-\sin^2\theta)\left[-\frac{R^2}{2M^2}dV^2+2dV dR+2M^2d\theta^2\right]\nonumber \\+\frac{4M^4\sin^2\theta}{2M^2-M^2\sin^2\theta}(d\Phi+\frac{R}{2M^2}dV)^2 \,,
\end{align}
where $\Phi=\phi^{\prime}-\log(R/\sqrt{2}M),$ and $t^{\prime}=V-2M/R$. Here we have set $M=|a|$ for the extreme case. This metric also played an important role in exploring the emergence of conformal symmetries near the horizon of a highly spinning black hole \cite{Guica:2008mu, Porfyriadis:2014fja}. After supertranslating this metric, we get \cite{Galoppo:2024vww}:
\begin{align}\label{kerrnh}ds^2=g_{VV}dV^2+g_{VR}dV dR+g_{V\Phi}dV d\Phi +g_{V\theta}dV d\theta \nonumber \\+g_{\theta\theta}d\theta^2+g_{\Phi\Phi}d\Phi^2 \,,
\end{align}
where, \[g_{\theta\theta}=(2M^2-M^2\sin^2\theta)\left(1-\frac{2}{M}\partial^2_{\theta}f\right)+M\cos2\theta \partial_{\theta}f \,,\] and \begin{align*}
    g_{\Phi\Phi} =\left[ \frac{4M}{(2-\sin^2\theta)}\right]&\times&\nonumber  \\
&\left[ M\sin^2 \theta-\frac{2\sin 2\theta}{1-\sin^2\theta}\partial_{\theta}f-\partial^2_{\Phi}f\right]\,.
\end{align*}
We do not need the explicit forms of the other metric components for the computation of the memory. The horizon of the supertranslated extreme-Kerr is at $R_h=M+\frac{1}{2}D^2f.$ As we know, a class of ISCOs (for prograde orbits) for an extreme Kerr black hole is situated almost at the horizon \cite{1972ApJ...178..347B}. Again, computing the shift in angular separation of the binary components situated close to the ISCO (say $r_0 = (1+0.001)R$), with ISCO at $R\sim M+\mathcal{O}(f)$, we get the following:
{\small\begin{align}\label{memory-kerr}
 & \Delta L\simeq -\frac{1}{4L_0} M^2 q_0 (12 \cos (2 \theta )+3 \cos (4 \theta )+1) \cos (2 \phi )\delta\theta^2\nonumber \\&+\frac{32 M^2 q_0 \sin ^2(\theta ) \cos (\theta ) \cos (2 \phi ) (5 \sin (\theta )+\sin (3 \theta )-2 \cos (\theta ))}{(\cos (2 \theta )+3)^2 L_0}\delta\phi^2.
\end{align}}\footnote{we have used $\phi$ instead of $\Phi$ for this expression. }
Here, we have considered $f=q f(\theta,\phi)=q {\rm Re}(Y_{22})$, where ${\rm Re}(Y_{22})$ is the real part of $l=m=2$ spherical harmonic $Y_{lm}(\theta,\phi)$.

\section{Frequency modulation of the binary }\label{freq-mod}
In the previous subsection, we argued that an inspiralling intermediate mass binary black hole (IMBBH) in the vicinity of horizon of supermassive black hole (SMBH) acts as a detector for detecting supertranslation (ST) memory. IMBBH loses energy and angular momentum to GWs which can be observed by LISA. The separation of the binary components $L= 2 a$ (we call it IMBBH separation) at any instant during inspiral can be related to GW frequency $f$ emitted at that time via Kepler's law as~\cite{Peters:1963ux,peters1964gravitational}:
\begin{equation*}
    a = \left[\frac{(1+z) M_{\mathrm{bin}}}{\pi^2 f^2}\right]^{1/3} \,,
\end{equation*}
where $a$ is the radius of the binary, $1+z= (1+z_c)(1+z_{\rm g}) $~accounts for the cosmological and gravitational redshifts of the IMBBH, and $M_{\mathrm{bin}}$ is the source-frame total mass of the IMBBH. Note that the gravitational redshift $z_g$ of GWs from IMBBH placed near the ISCO of central Schwarzschild SMBH is approximately $0.225$. Whereas, the GWs from IMBBH in the vicinity of horizon/ISCO of an extremal Kerr SMBH can either be redshifted or blueshifted depending on the direction of emission of GWs. The net redshift (here denoted by $z_{\rm g}$) is restricted to the range $[- 0.4,0.7]$. (See ~\cite{Igata:2019hkz} and Figure $6$ therein.)

Once the supertranslation `hair' is implanted on the central SMBH, the binary separation in IMBBH changes as a result of ST memory as given by Eqs.~(~\ref{memory-sch}, \ref{memory-kerr}). The sudden change in the IMBBH separation leads to a corresponding sudden change in the emitted GW frequency. This instantaneous frequency jump in the GW signal of the IMBBH should be observable. The detection of the ST memory can be quantified by calculating the mismatch between the original GW signal when there is no ST memory and the signal modulated due to the ST memory.

 GWs in general relativity have two polarizations i.e., plus $h_+$ and cross $h_{\times}$. For quasicircular binaries, the evolution of these polarizations is determined by the intrinsic source parameters ($\vec{\theta}$) i.e., component masses ($m_1$, $m_2$) and spins ($\vec{\chi}_1$, $\vec{\chi}_2$); and the extrinsic source parameters ($\vec{\lambda}_h$) i.e., luminosity distance $D_L$ to the source, inclination angle $\iota$ of the binary with respect to the observer's line of sight, time $t_c$ and phase $\phi_c$ at the coalescence. Note that instead of ($m_1$, $m_2$), it is more natural to use the chirpmass $\mathcal{M}_c = (m_1 m_2)^{3/5}/(m_1+ m_2)^{1/5}$ and mass ratio ${\bf q} = m_2/m_1$ \footnote{Here {\bf q} should not be confused with the supertranslation charge $q$ in Eqs.~(\ref{memory1},\ref{memory-sch}).} in the parameter estimation as the shape of gravitational waveform is more sensitive to the chirpmass~\cite{Cutler:1994ys}.  An interferometric GW detector, such as LISA, records the time-domain strain $h(t; \vec{\theta}, \vec{\lambda})$ due to an incoming GW signal, which can be expressed as a linear combination of the plus ($h_+$) and cross ($h_{\times}$) polarizations of GWs weighted by the respective antenna responses ($F_{+}, F_{\times}$):
%
%
\begin{equation}
\label{strain0}
    h_{a}(t; \vec{\theta}, \vec{\lambda}) = F_{+}(\vec{\lambda}_F)h_{+}(t; \vec{\theta}, \vec{\lambda}_h) + F_{\times}(\vec{\lambda}_F)h_{\times}(t; \vec{\theta}, \vec{\lambda}_h)\,,
\end{equation}
where $t$ is the detector-frame time, $\vec{\theta} = \lbrace{\mathcal{M}_c, {\bf q}, \vec{\chi}_1, \vec{\chi}_2 \rbrace}$, $\vec{\lambda}_F = \lbrace{\alpha, \delta, \psi \rbrace}$, $\vec{\lambda}_h = \lbrace{D_L, \iota, t_c, \phi_c \rbrace}$, and $\vec{\lambda} = \vec{\lambda}_F \cup \vec{\lambda}_h$.  $F_{+,\times}(\vec{\lambda}_F)$ are the antenna pattern functions determined by the extrinsic source parameters $\vec{\lambda}_F$ describing the sky-position of the source and polarization angle of the emitted GWs.  Sky position of the binary source with respect to the detector is described by right ascension $\alpha$ and declination $\delta$; the polarization angle of the incoming GW is denoted by $\psi$\footnote{Ideally antenna pattern functions are time dependent due to the orbital motion of LISA around the Sun. Since we are using sky-polarization-averaged antenna pattern functions and assume the sources to be isotropically distributed on the sky, this averaging will be the same with respect to different positions of LISA in its orbit.}.

Implantation of ST `hair' on the central SMBH leads to a sudden change in the IMBBH separation and a corresponding sudden change in the GW frequency emitted by the binary. This is equivalent to time-shifting the time-domain strain given by Eq.~\eqref{strain0} by some amount $\Delta t$ as:
\begin{align}\label{strain-shifted}
    h_{b}(t; \vec{\theta}, \vec{\lambda})=h_{a}(t+ \Delta t; \vec{\theta}, \vec{\lambda}) \, .
\end{align} 
In the {\tt stationary phase approximation} (SPA), the Fourier-domain expressions corresponding to Eqs.~\eqref{strain0}-\eqref{strain-shifted} can be expressed as~\cite{Droz:1999qx}:
\begin{align}\label{waveform}
    \Tilde{h}_{a}(f) = \mathcal{A} e^{i\Psi(f)} =\hat{\mathcal{A}} f^{-7/6} e^{i \Psi(f)} \,, \\ \nonumber
     \Tilde{h}_{b}(f)= e^{i 2\pi f \Delta t}  \Tilde{h}_{a}(f) \,.
\end{align}
Averaging over the inclination angle $\iota$, the amplitude parameter $\hat{\mathcal A}$ in the quadrupole approximation is given by~\cite{Moore:2016qxz,Favata:2021vhw}:
\begin{equation}\label{amplitude}
  \hat{\mathcal{A}} = \sqrt{\frac{5}{24}} \sqrt{\frac{4}{5}} \frac{\sqrt{\eta} M_{\rm bin}^{5/6} (1+z)^{5/6}}{\pi^{2/3} D_{L}} \,.
\end{equation}
Amplitude contains only inclination angle averaging factor ($\sqrt{4/5}$), while the factors due to $60^\circ$ angle between LISA arms ($(\sin^{2}60^\circ)^{-1}$) and  sky-polarisation-averaging ($(1/5)^{-1}$) are absorbed into the instrumental noise power spectral density (PSD) of the detector (discussed later in next section). In Eq.~(\eqref{amplitude}), $M_{\mathrm{bin}}=m_1+m_2$ is the total binary mass in the source frame with component masses $m_1$ and $m_2$ , $\eta=m_1 m_2/(m_1+m_2)^2= {\bf q}/(1+{\bf q})^2$ is the symmetric mass ratio. Assuming a flat universe, the luminosity distance-redshift relation is given as~\cite{hogg}:
\begin{equation}
\label{eq:dLz}
D_L(z) = \frac{c}{H_0} (1+z) \int_0^z \frac{dz'}{\sqrt{\Omega_M (1+z')^3 + \Omega_{\Lambda}}},
\end{equation} 
with the following cosmological parameters~\cite{Planck:2015fie}:
{$H_{0}=67.90 \,(\rm{km/s})/\rm{Mpc}$}, $\Omega_{M}=0.3065$, and $\Omega_{\Lambda}=0.6935$.

{\tt TaylorF2} SPA phase $\Psi(f)$ for circular binaries in the post-Newtonian (PN) theory is  expressed as a series in the orbital velocity parameter $v = [\pi M_{\mathrm{bin}}  (1+z) f]^{1/3}$ 
as~\cite{Blanchet:1995ez,Blanchet:1995fg,Kidder:1995zr,Blanchet:2002av,Blanchet:2006gy,Arun:2008kb,Marsat:2012fn,Mishra:2016whh}:
\begin{equation}
\label{phase}
     \Psi(f)  =  2\pi ft_{c} + \phi_{c} + \frac{3}{128 \eta v^{5}} \sum_{k} (\varphi_{k}v^{k} 
      +  \varphi_k^{\rm log}  v^{k} \ln v) \,,
\end{equation}
where $t_{c}$ and $\phi_{c}$ are the time and phase of coalescence, respectively. Corrections of order $v^{k}$ correspond to $k/2$-PN order correction relative to the first (0PN/Newtonian) term. The 3.5PN accurate expressions for $(\varphi_k, \varphi_k^{\, log})$ in terms of intrinsic source parameters (masses and spins) are taken from equations in 
Refs.~\cite{Arun:2004hn,Arun:2008kb,Buonanno:2009zt,Wade:2013hoa,Mishra:2016whh,Blanchet:2023bwj,Blanchet:2023sbv}. We consider the binary in quasicircular orbit and components to be nonspinning. 

Time-shift $\Delta t$ in Eq.~\eqref{waveform} corresponds to the frequency change due to supertranslation memory. Using the energy-balance equation,
adiabatic approximation and Kepler’s law, the expression
for  $\Delta t$ can be expressed at the 0PN (Newtonian) accuracy as~\cite{Buonanno:2009zt,Arun:2008kb,Arun:2009mc,Moore:2016qxz}:
\begin{align}\label{time-shift}
    \Delta t &=t(f_{\rm{fi}})-t(f_{\rm{in}}) \nonumber \\
    &= \frac{5}{256} ((1+z) \mathcal{M}_{c})^{-5/3} \Big(\frac{1}{(\pi f_{\rm{in}})^{8/3}}- \frac{1}{(\pi f_{\rm{fi}})^{8/3}}\Big)\,,
\end{align}
where $t(f)$ is the time corresponding to GW frequency $f$, $f_{\rm in}$ and $f_{\rm fi}$ are, respectively, the  GW frequencies at which binary is emitting before and after supertranslation hair is implanted on the central SMBH and ${\rm \mathcal{M}_{c}}= M_{\rm bin} \eta^{3/5}$ is the chirp mass of the binary. Eq.~\eqref{time-shift} can be expressed in terms of IMBBH mass $M_{\rm bin}$, initial IMBBH separation $L_0$ and upto the linear order in dimensionless supertranslation charge $q_0$ as
\begin{equation}\label{dt-schematic}
    \Delta t \propto  \frac{q_0 }{f_{\rm ref}^{8/3}\eta(1 + z_{\rm g})^{5/3}(1 + z_{c})^{5/3} M_{\rm bin}^{5/3}} \,,
\end{equation} where $z_{\rm g}$ is the gravitational redshift of gravitational waves emitted by IMBBH in the neighbourhood of SMBH. Note that in order to simplify that calculations, we assume the frequency jump due to ST memory from $f_{\rm  in}$ to $f_{\rm  fi}$ to be discrete. This corresponds to the maximum time shift in the IMBBH GW signal throughout the shock wave pulse and henceforth, the Bayes Factors obtained with this assumption will be considered as the upper limit. 

\section{Bayesian Parameter Inference}\label{bayesian-inference}
 Gravitational wave detector output is the time-domain strain data $d(t)$ which can be decomposed into two components i.e., the detector noise $n(t)$ and the GW signal strain $h(t)$. Given the waveform template $h(t; \vec{\theta}, \vec{\lambda})$  representing the GW signal, one needs to estimate the parameters $( \vec{\theta}, \vec{\lambda})$  of the GW source. Bayesian parameter estimation invokes Bayes Theorem to estimate the posterior probability distribution $p(\vec{\theta},\vec{\lambda}|d)$ of the signal parameters ($\vec{\theta}$, $\vec{\lambda}$) given the data $d$ and signal model $h(\vec{\theta}, \vec{\lambda})$ as:
\begin{align}\label{bayes-theorem}
 p(\vec{\theta}, \vec{\lambda}|d) = \frac{\Pi(\vec{\theta}, \vec{\lambda}) p(d| \vec{\theta}, \vec{\lambda})}{p(d)}\, ,
\end{align}
where $\Pi(\vec{\theta}, \vec{\lambda})$ represents the prior probability distribution, $p(d| \vec{\theta}, \vec{\lambda})$ represents the likelihood of the data $d$ and $p(d)$ is the evidence. $p(d)$ can be obtained by marginalizing the likelihood over the signal parameters as:
\begin{align}
    p(d) = \int \Pi(\vec{\theta}, \vec{\lambda}) p(d| \vec{\theta}, \vec{\lambda}) d\vec{\theta} d\vec{\lambda}\,.
\end{align}
Assuming the detector noise to be stationary and Gaussian, the likelihood can be written as~\cite{Cutler:1994ys}:
\begin{align*}
  p(d| \vec{\theta}, \vec{\lambda}) \propto e^{-\frac{1}{2} \langle d-h(\vec{\theta}, \vec{\lambda}) | d-h(\vec{\theta}, \vec{\lambda}) \rangle}\, ,  
\end{align*}
where $d$ is the strain data of the detector and $\langle .  |  . \rangle$ is the noise-weighted inner product defined as:
\begin{align}
\label{inner-product}
  \langle A | B \rangle  = 2 \int ^{\rm{f_{high}}}_{\rm{f_{low}}} \frac{A(f) B^{*}(f) + A^{*}(f) B(f)}{S_{n}(f)}df\,.
\end{align}
Here $f_{\rm low}$, $f_{\rm high}$ are determined by the properties of the binary source, observation time $T_{\rm obs}$ and the detector noise sensitivity. $S_{n}(f)$ is the noise power spectral density (PSD) of the detector. Note that the posteriors $p(\vec{\theta}, \vec{\lambda}|d)$ in Eq.~\eqref{bayes-theorem} are obtained by sampling the high-dimensional (usually fifteen-dimensional for quasicircular binaries) likelihood function approximated as Gaussian. 

Assuming the template $\tilde{h}(f)$ to accurately represent the GW signal and considering the detector noise to be stationary and Gaussian, the optimal signal-to-noise ratio (SNR) can expressed as:
\begin{align}
    \rho = \sqrt{4 \int^{\infty}_{0} \frac{\abs{\tilde{h}(f)}^{2} }{S_{n}(f)} df}\,.
\end{align}
Here noise PSD $S_{n}(f)$ of LISA consists of instrumental noise and confusion noise due to the galactic white dwarfs. Eq.~(1) of Ref.~\cite{babak2017science} provides the instrumental noise PSD and the confusion noise is taken from Eq.~(4) of Ref.~\cite{mangiagli2020observing}. The instrumental noise in Ref.~\cite{babak2017science} is sky-polarization-averaged and accounts for the $60^{\circ}$ angle between the arms of LISA. We divide the instrumental noise given in Ref.~\cite{babak2017science} by a factor of $2$ in order to account for the two effective L-shaped detectors within LISA.

The lower and upper cutoff frequencies ($f_{\rm low}$ and $f_{\rm high}$) used in calculating the integral in Eq.~\eqref{inner-product} are determined by the detector sensitivity, observation time, and the source properties.  The $f_{\rm low}$ is defined as $f_{\rm low}={\rm max}  [10^{-4},f_{\rm year}]\, {\rm Hz}$,
where $f_{\rm year}$ is the GW frequency ${\rm T_{obs}}$ years before the binary emits at the frequency corresponding to the innermost stable circular orbit (ISCO). This frequency, in the quadrupole radiation approximation, is determined by the chirp mass ${\mathcal M_{c}}$ and the observation time prior to ISCO $T_{\rm obs}$ as~\cite{berti2005estimating}:
\begin{equation}
f_{\rm year} = 4.149\times 10^{-5}\left(\frac{{\mathcal M_{c}}}{10^6 M_\odot}\right)^{-5/8}
\left(\frac{T_{\rm obs}}{1 \rm \text{year}}\right)^{-3/8} \,. 
\end{equation}
We assume $T_{\rm obs}=4$ year in our calculation. The upper cut-off frequency is given by $f_{\rm high} = {\rm min} [0.1,f_{\rm ISCO}]\, {\rm Hz}$,
where $f_{\rm ISCO} = 1/({6^{3/2}\pi (1+z) M_{\rm bin}})$ is the GW frequency corresponding to ISCO of a test particle orbiting around a  Schwarzschild black hole of mass $M_{\rm bin}$.

\begin{figure}[th]
   \centering 
 \includegraphics[width=\columnwidth]{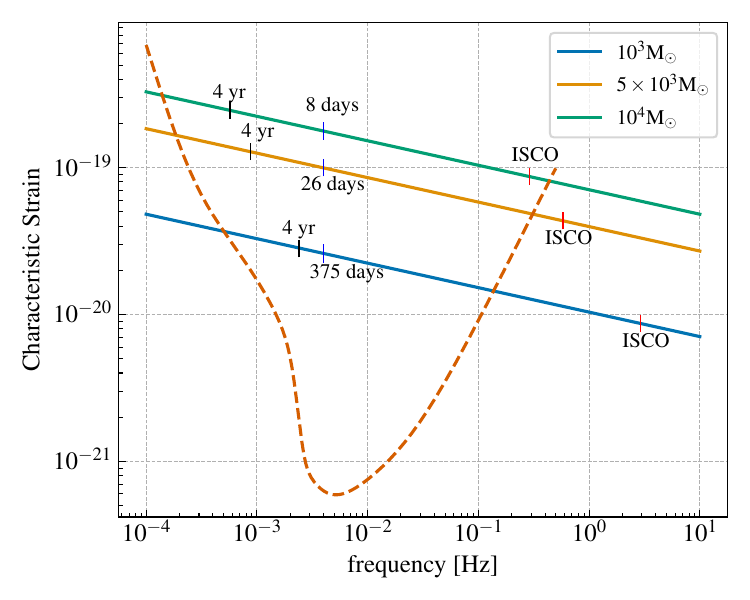}
  \caption{The evolution of the characteristic (dimensionless) strain~\cite{Moore:2014lga} (inclination-averaged) of three representative IMBBH sources with equal-mass components (solid lines) shown in comparison to the  characteristic (dimensionless) strain corresponding to the noise PSD of LISA (dashed curve). Luminosity distance of the IMBBH sources is fixed at $3$ Gpc. The red markers denote the GW strain and GW frequency of the IMBBH sources at the corresponding ISCO. The blue markers at $f_{\rm ref} = 4\, {\rm mHz}$ denote the time till ISCO. The black markers represent the characteristic strain, and GW frequency, $T_{\rm obs} = 4$ years before ISCO.} 
  \label{detector-strain}
\end{figure}

We have displayed in Fig.~\ref{detector-strain}, how IMBBH sources of different representative total masses ($10^3\, M_{\odot}$, $5 \times 10^3\,M_{\odot}$, $10^4\, M_{\odot}$) at a fixed luminosity distance of $3$ Gpc will evolve in the LISA band. Considering observation time $T_{\rm ob}=4\,{\rm year}$ till ISCO, these sources will inspiral in the LISA band starting from the frequencies corresponding to black markers, merge at frequencies corresponding to red markers. 

Let $f_{\rm ref}$ be the reference frequency to which the binary has evolved, when the ST hair is implanted on the SMBH. We choose this frequency (say, for example, $4\, {\rm mHz}$) within the higher sensitivity regime of LISA band. Indeed, the binaries are still $375$ days, $26$ days and $8$ days (shown by blue markers), respectively, away from reaching their respective ISCOs.

\section{Model selection}\label{bayes-factor}
In the context of this work, the GW signal emitted by the IMBBH can be modelled in two ways $\mbox{--}$ with ($\tilde{h}_{b}(f)$) and without ($\tilde{h}_{a}(f)$) ST memory. In order to determine whether one model (say $\tilde{h}_{b}(f)$) is preferred over another model (say $\tilde{h}_{b}(f)$), we need to do a comparative analysis which involves the calculation of Bayes factor. Bayes factor is defined as the ratio of the two marginalized likelihoods (evidences) calculated under each of the two hypotheses $\mbox{--}$ the hypothesis $\mathcal{H}_{b}$($\mathcal{H}_{a}$) that the GW signal in the data $d$ contains (doesn't contain) the supertranslation (ST) memory imprint. This factor can be expressed as: 
\begin{align}
    \mathcal{B}^{b}_{a} = \frac{p(d | \mathcal{H}_{b} )} {p(d | \mathcal{H}_{a})} = \frac{\int p(\vec{\theta}| \mathcal{H}_{b}) p(d | \vec{\theta}, \mathcal{H}_{b} ) d\vec{\theta}} {\int p(\vec{\theta}| \mathcal{H}_{a}) p(d | \vec{\theta}, \mathcal{H}_{a} ) d\vec{\theta}}\,.
\end{align}
 In the limit of high SNR, the above expression for Bayes factor can be simplified and written in terms of optimal SNR and fitting factor (FF) as~\cite{Cornish:2011ys,Vallisneri:2012qq}:
\begin{align}\label{BF-SNR}
    \text{ln} \mathcal{B}^{b}_{a} \approx (1-\text{FF}) \rho^2\,,
\end{align}
where ${\rm FF}$ is defined as the match $\mathcal{M}$ between the two waveform models $\tilde{h}_{b}(f)$ ($\tilde{h}_{a}(f)$) under two different hypotheses $\mathcal{H}_{b}$ ($\mathcal{H}_{a}$) maximized over the intrinsic source parameters $\vec{\theta}$ and can be written as~\cite{PhysRevD.52.605,Vallisneri:2012qq}:
\begin{align}
    \text {FF} \equiv \mathcal{M}_{ \text{max}(\vec{\theta})}\,. 
\end{align} 
The match $\mathcal{M}$ between two waveform models $\tilde{h}_{a}(f)$ and $\tilde{h}_{b}(f)$ quantifies the overlap between the two models and is defined as:
\begin{align}\label{match}
    \mathcal{M} = \frac{ \langle \tilde{h}_{a}(f)|\tilde{h}_{b}(f) \rangle   }{( \langle \tilde{h}_{a}(f)|\tilde{h}_{a}(f) \rangle  \langle \tilde{h}_{b}(f)|\tilde{h}_{b}(f) \rangle )^{1/2}}\,.
\end{align}
The denominator of the above expression can be simplified using Eq.~\eqref{waveform} as:
\begin{align}\label{overlap}
    \langle \tilde{h}_{a}(f)|\tilde{h}_{a}(f) \rangle = \langle \tilde{h}_{b}(f)|\tilde{h}_{b}(f) \rangle = \abs{\tilde{h}_{a}(f)}^{2}\,.
\end{align}
Using Eq.~\eqref{waveform}, $\tilde{h}_{a}(f)=\tilde{h}_{b}(f)$ only for $f < f_{\rm ref}$ just before the ST hair is implanted on the central SMBH (i.e., $\Delta t = 0$). 
Hence, using Eq.~\eqref{inner-product}, the numerator of Eq.~\eqref{match} can be written as a sum of two parts corresponding to $f < f_{\rm ref}$ and $f > f_{\rm ref}$ as:
\begin{align}\label{match-final}
   \langle \tilde{h}_{a}(f)|\tilde{h}_{b}(f) \rangle &=  4 \int^{f_{\rm ref}}_{f_{\rm low}} \frac{\abs{\tilde{h}_{a}(f)}^{2} }{S_{n}(f)} df  \\ \nonumber
   &+ 2 \int^{f_{\rm high}}_{f_{\rm ref}} \frac{2\abs{\tilde{h}_{a}(f)}^{2} \cos({2 \pi f \Delta t}) }{S_{n}(f)} df
\end{align}
Here $f_{\rm{ref}}$ is the GW frequency at which the binary is emitting corresponding to the time the ST memory is imparted to the binary.

\section{Results}
    \begin{figure*}[htb]
   \centering 
 \includegraphics[width=\textwidth]{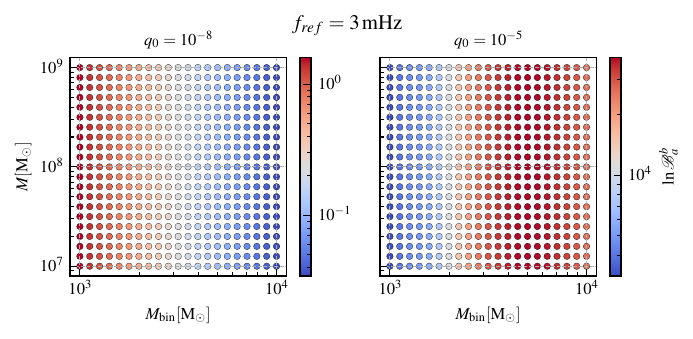}
 \includegraphics[width=\textwidth]{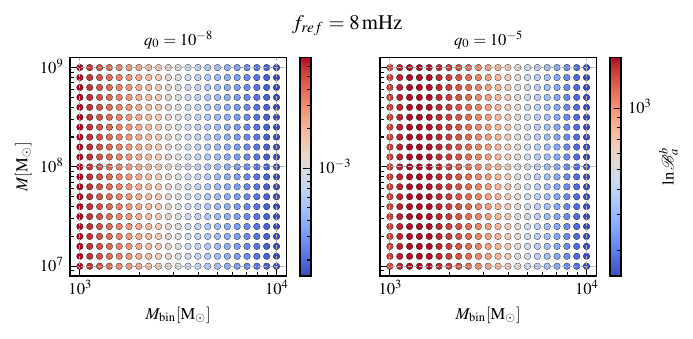}
  \caption{The {\tt Log-Bayes-Factor} $({\rm ln \mathcal{B}}^{\rm b}_{\rm a})$ to determine whether the model of GWs with (b)/without (a) imprints of black hole memory for Schwarzschild black hole is preferred, is shown in the color scale. These Bayes factors are evaluated on a grid of total binary mass ($M_{\rm bin}$) and mass of supermassive black hole ($M$). GW frequency corresponding to the time when supertranslation hair is imparted to SMBH is fixed at $f_{\rm ref}=3\, {\rm mHz}$ (top row) or $f_{\rm ref}=8\, {\rm mHz}$ (bottom row). Dimensionless supertranslation charge $q_0$ is varied from left ($q_0= 10^{-8}$) to right ($q_0= 10^{-5}$) panels as shown. This figure corresponds to the case when ST memory leads to an increase (corresponding to $\cos(2\theta)= -1$ in Eq.~\eqref{memory-sch}) in binary separation.} 
  \label{BF-positive-fref-fixed-sch}
\end{figure*} 
 
\begin{figure*}[htb]
   \centering 
 \includegraphics[width=\textwidth]{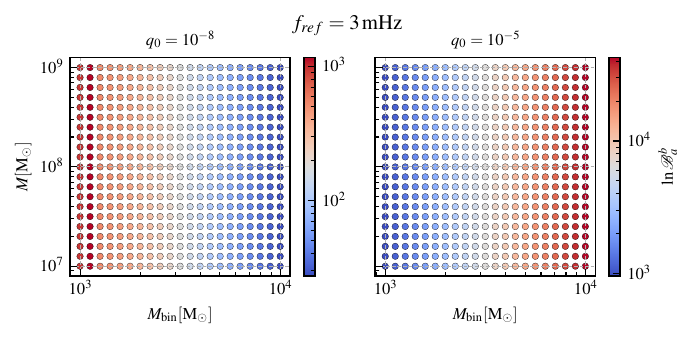}
 \includegraphics[width=\textwidth]{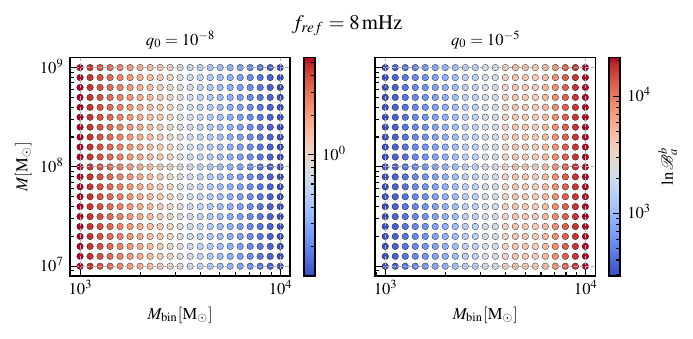}
  \caption{ Same as Figure~\ref{BF-positive-fref-fixed-sch} but for extremal Kerr black hole (redshift $z_g=-0.4$) and this Figure corresponds to the case when ST memory leads to an increase (corresponding to $\cos(2\theta)= 0$ and $\cos(\phi)= 1$ in Eq.~\eqref{memory-kerr} ) in binary separation.} 
  \label{BF-positive-fref-fixed-kerr-minus}
\end{figure*} 

\begin{figure*}[htb]
   \centering 
 \includegraphics[width=\textwidth]{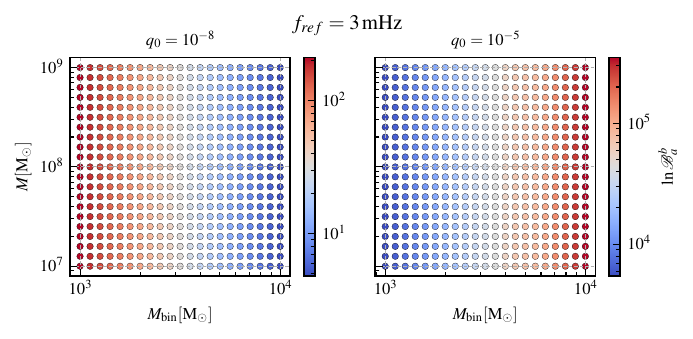}
 \includegraphics[width=\textwidth]{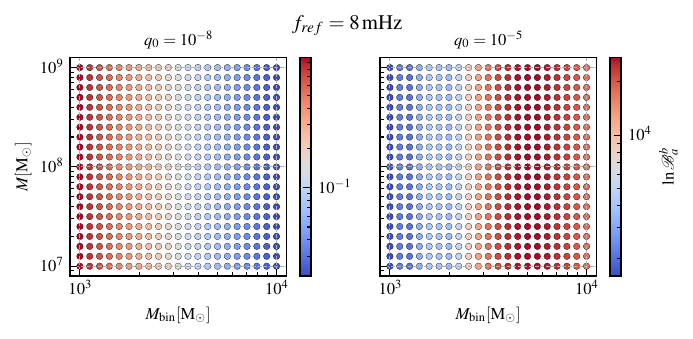}
  \caption{ Same as Figure~\ref{BF-positive-fref-fixed-sch} but for extremal Kerr black hole (redshift $z_g=0.7$) and this Figure corresponds to the case when ST memory leads to an increase (corresponding to $\cos(2\theta)= 0$ and $\cos(\phi)= 1$ in Eq.~\eqref{memory-kerr} ) in binary separation.  } 
  \label{BF-positive-fref-varied-kerr-plus}
\end{figure*} 

Having introduced the waveform models in Sec.~\ref{freq-mod}, Bayesian inference in Sec.~\ref{bayesian-inference} and model comparison in Sec.~\ref{bayes-factor}, we discuss the results here.
We construct a $21 \times 21$ grid {\tt uniform-in-log} of the total mass $M_{\rm{bin}}$  of IMBBH and  {\tt uniform-in-log} of the mass $M$ of SMBH.  The masses of the IMBBH and SMBH are restricted to the ranges $10^3 \mbox{--} 10^4\, \rm{M_\odot}$ and $10^7 \mbox{--} 10^9\, \rm{M_\odot}$, respectively. 
We fix the mass-ratio at {\bf q} $= 1$ (corresponding to $m_1=m_2$). The binary components of the IMBBH source are assumed to be nonspinning. The luminosity distance $D_L$ of the IMBBH source (and SMBH as well) is chosen to be $3$ Gpc. We calculate the SNRs of the simulated sources within the LISA band and find that the SNRs fall in the ranges ($50-400$), ($30-220$) and ($75-520$) for Swarzschild case, extreme Kerr cases with redshifts $z_{\rm g}=-0.4$ and $z_{\rm g}=0.7$, respectively. These SNRs are high enough to validate the high-SNR approximation used in Eq.~\eqref{BF-SNR} of Sec.~\ref{bayes-factor} for simplifying the calculation of the Bayes factor~\cite{Vallisneri:2007ev,2008CQGra..25r4007C,Rodriguez:2013mla}.

The supertranslation charge $q_0$ appearing in Eqs.~(\ref{memory-sch},\ref{memory-kerr}) is chosen from a representative set of discrete values: ($10^{-8}$, $10^{-5}$) and the reference GW frequency $f_{\rm{ref}}$, corresponding to the time at which ST memory is imparted to the binary, is chosen to be ($3\, \rm{mHz}$, $8\, \rm{mHz}$). This corresponds to the significant fraction of bottom of the noise PSD ``bucket'' (Fig.~\ref{detector-strain}) where the LISA detector is the most sensitive.
 
To show that the GW signal prefers the model with ST memory over the model without ST memory, we calculate Bayes-factor for all the simulated sources using the parameter setup discussed. Figure~\ref{BF-positive-fref-fixed-sch} (corresponding to $\cos 2\theta = -1$ in Eq.~\eqref{memory-sch}) shows the scatter-plot for {\tt Log-Bayes-Factor} in the $M-M_{\rm{bin}}$ plane. The different panels in each row of Fig.~\ref{BF-positive-fref-fixed-sch} pertain to different values of dimensionless supertranslation charge $q_0$ while keeping the reference frequency fixed as mentioned. We see that {\tt Log-Bayes-Factor} shows a non-trivial trend with the total mass $M_{\rm{bin}}$ of the IMBBH. This is because {\tt Log-Bayes-Factor} as given in Eq.~\eqref{BF-SNR} depends on two factors, fitting factor $FF$ [see Eqs. (~\ref{match}-~\ref{match-final})] and square of the SNR $\rho^2$. $FF$ has a sinusoidal dependence on time-shift $\Delta t$ with $\Delta t$ having a decreasing trend with increase in the binary mass $M_{\rm bin}$ [see Eq.~\eqref{dt-schematic}] and SNR $\rho$ increases with increase in mass of the binary $M_{\rm bin}$. These two competing factors decide the net dependence of {\tt Log-Bayes-Factor} on the binary mass $M_{\rm bin}$ which also depends on the combination of other parameters like dimensionless ST charge $q_0$, reference frequency $f_{\rm ref}$, gravitational/relativistic redshift/blueshift $z_{\rm g}$ and symmetric mass ratio $\eta$. Given a total mass $M_{\rm{bin}}$ of the binary, {\tt Log-Bayes-Factor} doesn't shows any dependence on mass $M$ of SMBH. This is because $\delta \phi$ and $\delta \theta$ in Eqs. (\ref{memory-sch},\ref{memory-kerr}) are inversely proportional to $M^2$ making $\Delta L$ and hence $\Delta t$ [Eq.~\eqref{dt-schematic}] independent of M. Increasing the value of ST charge $q_0$ increases the strength of supertranslation memory as shown in different panels of each row of Fig.~\ref{BF-positive-fref-fixed-sch}. When the ST charge $q_0$ is higher, we get higher {\tt Log-Bayes-Factor} for given masses $M_{\rm{bin}}$, $M$ of the binary and SMBH, respectively, as $\Delta L \propto q_0$. Moreover, as we move from top row ($f_{\rm ref}=3\, {\rm mHz}$) to bottom row ($f_{\rm ref}=8\, {\rm mHz}$) of Fig.~\ref{BF-positive-fref-fixed-sch}, {\tt Log-Bayes-Factor} reduces by nearly two orders of magnitude. This is because $\delta \phi$ and $\delta \theta$ in Eq.~\eqref{memory-sch} are directly proportional to $L_0^2$ and hence making $\Delta L$ proportional to $L_0$ ($L_0 \propto f_{\rm ref}^{-2/3}$).

Figures~(\ref{BF-positive-fref-fixed-kerr-minus}, \ref{BF-positive-fref-varied-kerr-plus}) correspond to the case when IMBBH is nearly at the ISCO/horizon ($r \approx \, (1+ 0.001) M$) of an extremal Kerr SMBH with extreme redshift choices of $z_{\rm g}= -0.4$ and $z_{\rm g}= 0.7$, respectively, from the allowed redshift range $z_{\rm g} \in [-0.4,0.7]$~\cite{Igata:2019hkz}. The trends in the {\tt Log-Bayes-Factor} with respect to dimensionless ST charge $q_0$, reference frequency $f_{\rm ref}$, IMBBH mass $M_{\rm bin}$ and SMBH mass $M$ hold the similar explanations as for Fig.~\ref{BF-positive-fref-fixed-sch}. Comparing the {\tt Log-Bayes-Factor} in Fig.~\ref{BF-positive-fref-fixed-kerr-minus} with the correspondings numbers in Fig.~\ref{BF-positive-fref-fixed-sch}, there is nearly three(one) order(s) of improvement across the chosen values of reference frequency $f_{\rm ref}$ with fixed $q_{0} = 10^{-8}$ ($10^{-5}$). Similarly, comparing the {\tt Log-Bayes-Factor} in Fig.~\ref{BF-positive-fref-varied-kerr-plus} with the corresponding numbers in Fig.~\ref{BF-positive-fref-fixed-sch}, there is approximately two(one) orders(s) of improvement across the chosen values of reference frequency $f_{\rm ref}$ with fixed $q_{0} = 10^{-8}$ ($10^{-5}$). This suggests that the detection prospects of black hole memory in the vicinity of the horizon of extremal Kerr SMBH are better compared to corresponding detection probability near ISCO of Schwarzschild SMBH. Moreover, it is important to note that we also computed the {\tt Log-Bayes-Factor} for the Kerr Case with intermediate values of  redshift $z_{\rm g}=0.2, 0.5$ and we didn't notice any drastic change in the results when comapred to the Kerr case with redshift $z_{\rm g}=0.7$ [corresponding to Fig.~\ref{BF-positive-fref-varied-kerr-plus}].  

For all chosen values of dimensionless ST charge $q_0$ and reference frequency $f_{\rm ref}$, {\tt Log-Bayes-Factor} is the best for a larger supertranslation charge ($q_0=10^{-5}$)  and a smaller reference frequency ($f_{\rm ref}=3\, {\rm mHz}$) for both the Schwarzschild ($\ln \mathcal{B}^{b}_{a} \approx 10^3-10^4$) and extremal Kerr ($\ln \mathcal{B}^{b}_{a} \approx 10^4-10^5$) SMBHs. This is because $\Delta t$ in Eq.~\eqref{dt-schematic} is directly proportional to $q_0$ and the dependence on reference frequency $f_{\rm ref}$ is demonstrated via $\Delta t \propto f^{-8/3}_{\rm ref}$. The large {\tt Log-Bayes-Factor} obtained for a significant portion of the parameter space considered, for both Schwarzchild and extremal Kerr SMBHs,  show that the near-horizon ST memory can be confidently detected in the LISA band. 
  
\section{Conclusion}
We have presented, as a proof of principle, two models to detect the horizon supertranslation memory. The first set up involves an IMBBH emitting GWs detectable by LISA, in the vicinity of a non-rotating SMBH suddenly endowed with the memory, which acts as a proxy for a detector. The sudden change in the spacetime metric surrounding the SMBH produces a sharp change in the frequency and frequency evolution of the GWs, which can be exploited to identify the occurrence of this memory. We further considered another case where the SMBH is an extreme Kerr black hole and determined the frequency change of the GWs. We refer the reader to \cite{goncharov2024inferring} for other possible mechanisms that could enable the detection of BMS-symmetries through GW memory in next generation (XG) detectors. 

The detection scheme outlined here, as a proof of principle, possesses caveats that would need to be addressed to realistically detect the black hole memory. 
\begin{itemize}
    \item The source of the shockwave in a realistic astrophysical scenario is unknown, although many interesting results have been reported in theoretical studies of asymptotic symmetries and memory effects using this model~\cite{PhysRevLett.116.231301,Hawking2017,strominger2018lectures,PhysRevLett.116.091101,PhysRevD.98.124016,Bhattacharjee2021,Sarkar:2021djs,Donnay2016,Chu:2018tzu}. 
    \item The GW waveform used for the non-spinning IMBBH does not account for environmental effects. The presence of the SMBH would modulate the waveform due to effects such as line of sight acceleration (see, e.g., \cite{Vijaykumar:2023tjg}), aberration (see, e.g., \cite{aberration}) and lensing (see, e.g., \cite{lensing}). In fact, to fully capture all the effects of SMBH on the GWs will require ray-tracing. Nevertheless, we expect the sudden change in the metric producing a discrete jump in the frequency of the IMBBH's GWs to be detectable over and above all the additional modulations incurred due to the presence of the SMBH.
    \item The rate of IMBBH mergers, or for that matter, CBCs in general in the vicinity of SMBHs, remains uncertain. Some works in the literature have proposed the existence of migration traps in disks of active galactic nuclei (AGNs). These traps could foster CBC mergers, some of which could lie very close to the ISCO of the central SMBH (see, e.g., \cite{migrationtraps}).
\end{itemize}
Another plausible mechanism to detect the black hole memory involves extreme mass ratio inspirals (EMRIs). In this case, a stellar mass black hole orbits an SMBH, which is suddenly endowed with a ``radial'' (instead of ``azimuthal'' as considered in this work) ST hair, causing the orbital radius of the EMRI to suddenly increase, thus modulating it. We expect the signature of such an ST memory also to be detectable. We plan to calculate this radial ST hair and the corresponding change in the metric, as well as its signature on the EMRI GWs, to provide some insights on this alternative detection setup.

\section*{Acknowledgments}
We thank K. G. Arun and Swastik Bhattacharya for their feedback on the manuscript. S. A. B. acknowledges useful comments from Sourabh Magare and Avinash Tiwari. S.B. acknowledges warm hospitality at IUCAA, Pune during which the work was initiated. S.B. thanks the participants of the International Conference of Gravity \& Cosmology (ICGC, 2023) held at IIT Guwahati for their inputs on an initial version of this work. S.A.B is grateful to the LISA Consortium participants for useful comments during the presentation of this work in the nascent stage as an invited talk in LISA Community Call and to the LISA Social Media Team (LSMT+) for advertising this work on LISA social media plateforms. The research of S. B. is supported by DST-SERB through MATRICS grant MTR/2022/000170. S.J.K. acknowledges support from SERB grants SRG/2023/000419 and MTR/2023/000086.  
\bibliographystyle{apsrev}
\bibliography{ref-list}
\end{document}